\documentclass[twocolumn]{IEEEtran}

\usepackage{tabularx} 
\usepackage{array} 
\usepackage{cite}
\usepackage{graphicx}
\usepackage{xcolor}
\usepackage{amsmath}
\usepackage{amssymb}
\usepackage{bm}
\usepackage{hyperref}
\hypersetup{
colorlinks,
citecolor=blue,
linkcolor=blue}
\usepackage{enumitem}
\usepackage{svg}
\usepackage{makecell}
\usepackage{multirow}
\usepackage{float}
\usepackage{caption}
\usepackage{subcaption}
\usepackage{soul}
\usepackage{booktabs}

\begin{document}

\title{Operational risk quantification of power grids using graph neural network surrogates of the DC OPF}

\author{Yadong Zhang \IEEEauthorrefmark{1}, Pranav Karve \IEEEauthorrefmark{1}, Sankaran Mahadevan \IEEEauthorrefmark{1} \IEEEauthorrefmark{2} 
\thanks{
\IEEEauthorblockA{\IEEEauthorrefmark{1}  School of Engineering, Vanderbilt University, Nashville, TN 37235, USA.}\\
\IEEEauthorblockA{\IEEEauthorrefmark{2} Corresponding author, E-mail: sankaran.mahadevan@vanderbilt.edu}
}}
%

\maketitle


\begin{abstract}
A DC OPF surrogate modeling framework is developed for Monte Carlo (MC) sampling-based risk quantification in power grid operation. MC simulation necessitates solving a large number of DC OPF problems corresponding to the samples of stochastic grid variables (power demand and renewable generation), which is computationally prohibitive. Computationally inexpensive surrogates of OPF provide an attractive alternative for expedited MC simulation. Graph neural network (GNN) surrogates of DC OPF, which are especially suitable to graph-structured data, are employed in this work. Previously developed DC OPF surrogate models have focused on accurate operational decision-making and not on risk quantification. Here, risk quantification-specific aspects of DC OPF surrogate evaluation is the main focus. To this end, the proposed GNN surrogates are evaluated using realistic joint probability distributions, quantification of their risk estimation accuracy, and investigation of their generalizability. Four synthetic grids (Case118, Case300, Case1354pegase, and Case2848rte) are used for surrogate model performance evaluation. It is shown that the GNN surrogates are sufficiently accurate for predicting the (bus-level, branch-level and system-level) grid state and enable fast as well as accurate operational risk quantification for power grids. The article thus develops tools for fast reliability and risk quantification in real-world power grids using GNN-based surrogates.
\end{abstract}

\begin{IEEEkeywords}
Risk assessment, system reliability, DC OPF surrogates, graph neural network
\end{IEEEkeywords}

\section{Introduction}

The growing participation of renewable energy resources (RES), bulk storage, natural gas turbines and other flexible generation sources (rooftop solar, vehicle to grid, etc.) leads to increasing volatility and uncertainty in the power grid~\cite{Milligan2010-yt,Rejc2014-dh,Heptonstall2020-sw}. Quantitative risk assessment and management tools are needed to safely operate the grid under these conditions, specifically: a) advanced decision-making algorithms that explicitly consider the uncertainty in grid variables, and b) explicit reliability and risk quantification methods. 

The first category includes stochastic optimization (i.e., stochastic programming, robust, chance-constrained, etc.)~\cite{Zheng2015-gu, Hong2021-tn} or dynamic reserve determination algorithms~\cite{Holttinen2013-gl, Mohandes2019-mm, Ela2010-ba}. These methods model the grid uncertainty using various uncertainty representation methods (scenarios, intervals, etc.) and allow operators to make unit commitment and dispatch decisions under uncertainty. Due to the high computational cost of solving these complex stochastic optimization problems, the grid uncertainty can only be approximately captured (using, e.g., only up to 50 scenarios) for real-world power grids. Grid security constraints are satisfied for the chosen (approximate) uncertainty representation, and the risk underlying the optimal solution (the decision) is never explicitly communicated. The risk quantification is thus implicit and approximate~\cite{Stover2023-ov}. Nevertheless, these methods are vital for optimal, safe, and reliable operation of the grid.

The second category of tools involves explicit, Monte Carlo (MC) sampling-based risk quantification given a commitment decision and a probabilistic forecast of grid variables~\cite{Stover2023-ov,Hong2021-tn,Cepin2011-pb}. These methods provide more accurate (\emph{high-resolution}) risk estimates than those used in the current decision-making algorithms, because they use numerous (thousands or more) MC samples. During day-to-day operations, the risk estimates obtained using these methods can improve grid operators' situational awareness for better decision-making. With heightened situational awareness, operators can tune the hyper-parameters (penalties, constraints, number of samples) of intra-day decision making algorithms (e.g., look-ahead commitment, or LAC) that ensure grid reliability. The real-time risk estimate provides a quantitative basis for emergency decisions such as purchasing additional power, asking consumers to shed loads etc.~\cite{dehghanian2016probabilistic}; such decisions are currently not based on rigorous risk quantification. The risk quantification can also enable risk versus cost analysis for competing methods from the first category to help arrive at the most suitable decision-making algorithm.

The main challenge in the adoption of sampling-based, explicit risk quantification for power grid operation is the computational cost of obtaining decisions (grid state) corresponding to numerous future scenarios of input variables~\cite{shafiullah2013potential}. This requires solving thousands of security-constrained (operational or cost) optimization problems such as the DC OPF problem~\cite{momoh1997challenges,dommel1968optimal,bakirtzis2003decentralized,biskas2005decentralized,xu2016real}. Using the state-of-the-art solution algorithms (e.g., alternating directional method of multiplier (ADMM)~\cite{wang2016fully,biagioni2020learning,yang2019distributed}), it is possible to solve DC OPF within 0.1~$s$ for small to medium size power grids~\cite{erseghe2014distributed}. However, it still takes up to a second to solve DC OPF for large grids ($>$~1000 buses)~\cite{abraham2018admm}. As a result, it may take more than thirty minutes to solve thousands of DC OPF problems to obtain samples of grid state corresponding to the samples of inputs. This is much longer than the typical decision-making window of five minutes for SCED (security-constrained economic dispatch). Therefore, it is necessary to develop fast surrogate models or optimization proxies that significantly ease the computational burden of sampling-based risk assessment.

The need for fast decision-making has led to significant research in developing machine learning-based methods that accelerate the OPF solution process~\cite{pineda2020data,deka2019learning,zhao2020deepopf+,velloso2021combining,chen2022learning,zhang2021convex,biagioni2020learning}. For instance, Deka and Misra~\cite{deka2019learning} leverage neural networks to learn active constraints that can maintain the optimality and achieve speedup  by reducing the problem size. Biagoni et al.~\cite{biagioni2020learning} developed RNN model that can predict the warm-start point to expedite the iteration process of the ADMM algorithm. These methods inevitably involve an iterative solution process and may increase the computational cost. To overcome this issue, end-to-end prediction (surrogate) models are also developed. These models are trained to learn the mapping between input (e.g., load) and output (generator dispatch)~\cite{dobbe2019toward,lei2020data,singh2021learning}. We refer the reader to~\cite{pan2022deepopf,zhao2020deepopf+,hasan2020survey} for further details of various end-to-end OPF surrogate models.

These previous efforts on OPF surrogate development focus on expediting the OPF solution for making operational decisions (setting power dispatch). When such models are to be used for risk assessment, the \emph{risk assessment surrogate} needs to be evaluated for various additional criteria not typically used in evaluating a \emph{decision-making surrogate}. These include: (a) using realistic marginal distribution types for the input variables; (b) considering the \emph{joint probability distribution} of, and hence the correlation between, input variables; (c) quantifying the accuracy of reliability/risk estimates and (d) investigating the generalizability of the model for test scenarios differing from the training data. Previously reported studies on OPF decision-making surrogates have not considered these criteria for model evaluation. In this article, we discuss a methodology for rigorously evaluating different surrogates developed specifically for enabling fast reliability/risk assessment.

The proposed methodology could be used to evaluate any end-to-end \emph{risk} surrogate model, including the ones previously reported in the literature. Here, we investigate graph neural network (GNN) surrogate models. GNN models are chosen because they can effectively deal with graph-structured data encountered in a power grid operation~\cite{zhang2023power,Bolz2019-zt,Yang2020-yb,Wang2020-wm,Xu2020-fr,Miao2020-du,James2019-jx}. Multiple GNN surrogates are trained using a supervised learning approach to predict quantities of interest (QoIs) at different levels: bus-level (active power output), branch-level (transmission line flow) and system-level (operating reserve, load shedding, and total cost). A joint probabilistic forecast is used to generate MC samples of power demand and renewable (wind) generation, which are provided as input to an OPF solver. Using the OPF input-output data for these samples, GNN surrogates are constructed for fast prediction of the bus-, branch-, and system-level QoIs. Predictions of QoIs are used for grid operational risk quantification using risk metrics relevant for day-to-day power grid operation. The reliability and risk estimate obtained using the OPF solver is considered as the ground truth, and the performance of the GNN surrogate is evaluated by comparing GNN-based reliability and risk estimates against OPF-based reliability and risk estimates. To evaluate the generalization capability, a distance metric is used to quantify the distance between the training and testing data distributions, and the effect of this distance on the accuracy of GNN-based risk assessment is quantified. The results provide insights into the GNN surrogate’s performance under different potential grid variable distributions (forecasts) and enhance the robustness of the decision-making process. The main contributions of this work are as follows:
\begin{enumerate}
    \item Investigation of the predictive capability of the DC OPF surrogate, i.e., GNN, for stochastic grid variable states drawn from realistic probabilistic forecasts.     This is the first effort to evaluate surrogate models using realistic (forecast) probability distributions.
    \item Evaluation of the DC OPF surrogate’s accuracy in estimating power grid operational reliability and risk. 
    \item Extension of a previously developed zonal and system-level reliability and risk assessment framework to quantify reliability and risk at the transmission line level. 
    \item Evaluation of the generalization capability of the DC OPF surrogates in the context of power grid operational risk quantification. Various forecasts of joint probability distributions of renewable generation and load variables are considered to mimic a wide range of operational scenarios. A statistical distance metric is used to measure how well the training data covers the forecast scenario, in order to evaluate the quality of the GNN model prediction for scenarios that significantly differ from the training data. The error in the GNN-based reliability and risk quantification is examined as a function of this distance.
\end{enumerate}

The remainder of this article is organized as follows. In  Section~\ref{sec:methodology}, the formulation of DC OPF is described and the methodology for GNN-based risk quantification is discussed. Numerical experiments are described in Section~\ref{sec:numerical_experiment} and the results are summarized in Section~\ref{sec:results}. Concluding remarks are provided in Section~\ref{sec:conclusion}.

\section{Methodology}
\label{sec:methodology}

In this section, we provide details of the DC OPF formulation and its GNN surrogates used in this work, the training data generation process, and method to evaluate the GNN surrogate model’s performance in estimating the grid reliability and risk.

\subsection{DC optimal power flow (OPF)}
\label{sec:dcopf}

The optimal power flow (OPF) problem is central to power system operation and plays a crucial role in economic dispatch, unit commitment, stability and reliability assessment, demand response, etc. The objective of the OPF problem is to minimize overall power generation cost while satisfying the grid security constraints. OPF in the full form (AC OPF) is nonlinear and nonconvex, which limits its usage in practice~\cite{Duan2018-dk,Pan2020-nk}. Typically, a simplified, linearized version of the OPF (DC OPF) is used. DC OPF ignores reactive power and voltage magnitude in the formulation. This guarantees convexity in the feasible domain and ensures that the solution exists~\cite{deka2019learning,bakirtzis2003decentralized}. DC OPF is routinely employed in electricity market clearing and power transmission management~\cite{Montoya2019-yh,bakirtzis2003decentralized,deka2019learning}. For this reason, DC OPF is the focus of this work. We use OPF to refer to DC OPF in the following text unless otherwise specified. The OPF formulation used in this work is given by:
\begin{align}
    \min \;\; & \sum_i c_i P_{G_i}, \nonumber \\
    \text{s.t.} \;\; & \sum_i P_{G_i} = \sum_i P_{D_i} \nonumber \\
    & P_{G_i}^{min} \leq P_{G_i} \leq P_{G_i}^{max} \\
    & \mathbf{B}\Theta = \mathbf{P}_G - \mathbf{P}_D \nonumber, \\
    & \frac{1}{x_{ij}} \left( \theta_i - \theta_j \right) \leq P_{i,j,max} \nonumber
\end{align}
\noindent where $c_i$ is marginal price of power generation, $P_{G_i}$ and $P_{D_i}$ denote power generation and demand, $P_{G_i}^{min}$ and $P_{G_i}^{max}$ denote the minimum and maximum allowed power generation at bus $i$, respectively, $\mathbf{B}$ and $\Theta$ represent admittance matrix and voltage angle vector, $x_{i, j}$,  $P_{i,j,max}$ are line reactance and line flow limit from bus $i$ to bus $j$, and $\theta_i$ is voltage angle for bus $i$.

\subsection{GNN surrogate development}

\begin{figure}
     \centering
     \includegraphics[width=\linewidth]{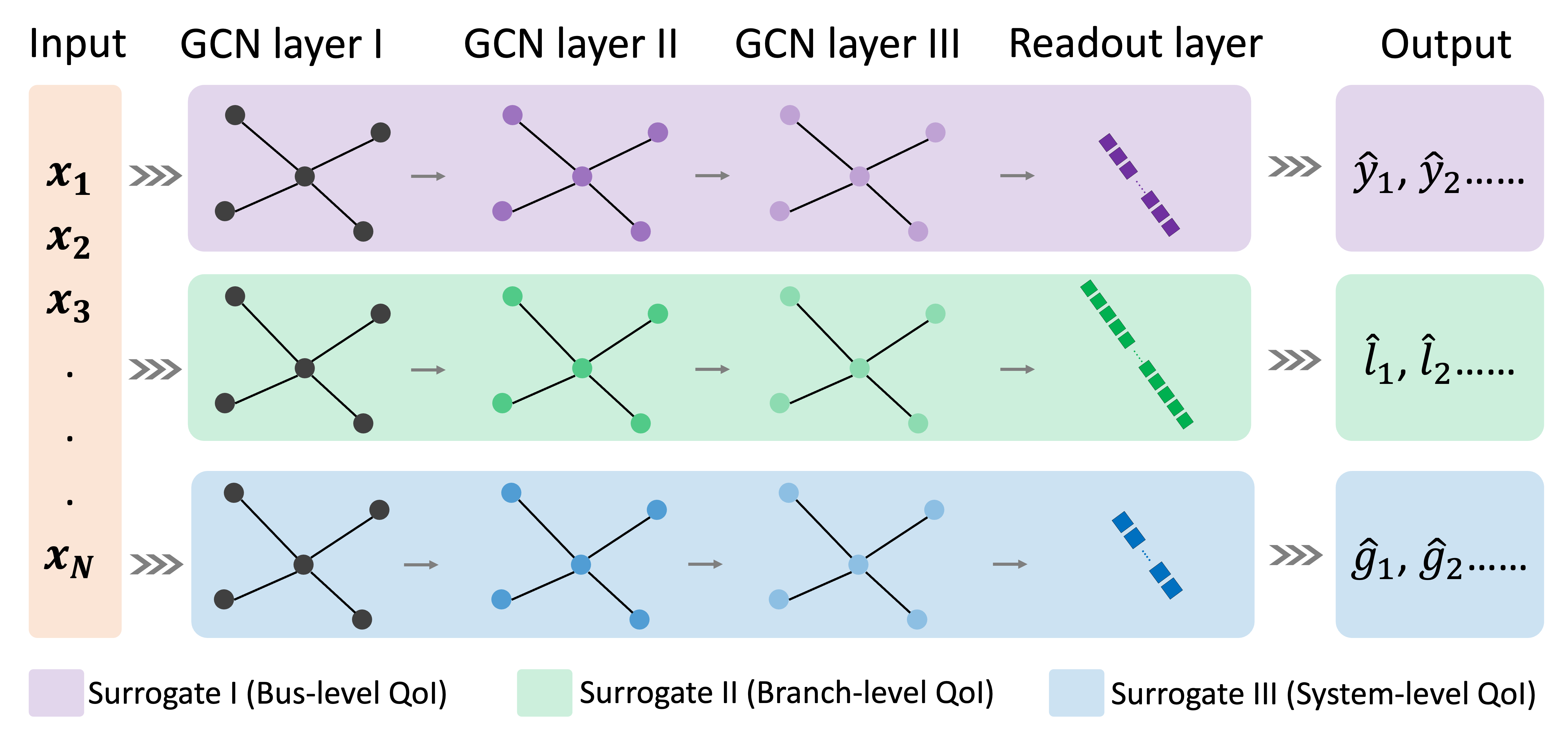}
     \caption{Structure of the three GNN surrogates.}
     \label{fig:GNN_demonstration}
\end{figure}

The GNN surrogates used in this article are constructed using three GCN layers followed by a single, fully connected readout layer. GCN layers are used to update the node representation while the readout layer is only used to output the QoI with the correct dimension. Since we consider QoIs at multiple levels (bus-, branch- and system-level), it is necessary to develop separate GNN surrogates. Three separate GNN surrogates are developed and each of them predicts the QoIs at one of the three levels.  The structure of these GNN surrogates are depicted in Fig.~\ref{fig:GNN_demonstration}. 
 
The loss function is constructed using the mean squared error (MSE) loss and a penalty term. The MSE loss is given by:
\begin{align}
    \label{MSE_loss}
    L_E = \frac{1}{|\mathcal{V}|}\sum_{i \in \mathcal{V}}(x_i-x_i^\ast)^2,
\end{align}
where $x_i^\ast$ and $x_i$ represent the training data value and predicted value, respectively. A penalty term is added to ensure that the predictions satisfy OPF inequality constraints. The penalty term is expressed as:
\begin{align}
    \label{auxiliary_variable}
    L_{IE} = \frac{1}{|\mathcal{V}|}\sum_{i \in \mathcal{V}}\zeta_i^2,
\end{align}
where $\zeta_i$ is an auxiliary variable defined as:
\begin{align}
    \label{penalty_term}
    \zeta_i = \text{max}\{x_i-x^{\rm max}_i, 0\} - \text{min}\{x-x^{\rm min}_i, 0\}
\end{align}
with $x^{\rm min}$ and $x^{\rm max}$ denoting the minimum and maximum allowed value of the QoI. Using weighted averaging, the final loss function can be represented as:
\begin{align}
    \label{final_loss}
    L = \omega_1L_E + \omega_2L_{IE}.
\end{align}
Here $\omega_1 > 0$ and $\omega_2 > 0$ are weights and $\omega_1 + \omega_2 = 1$. Equal weights ($\omega_1 = \omega_2 = 0.5$) could be used to give equal importance to both constraints. 

\subsection{Training data generation}
\label{sec:training_data_generation}

The stochastic variabilities in the power demand and renewable generation sources (supply) are the main sources of uncertainty in a power grid~\cite{Holttinen2013-iv,Mohandes2019-mm,Ela2010-ba,Zheng2015-gu}. The probability distribution that quantifies such variability is not known at the time of training data generation, since it is to be obtained through forecasting for a future time of interest. The training data for the GNN surrogate is therefore generated by considering uniform distributions for the demand $P^L_i, \; \forall \, i \in \mathcal{V}_L$ and renewable generation $P^R_i, \; \forall \, i \in \mathcal{V}_R$, where $\mathcal{V}_L$ and $\mathcal{V}_R$ denote the set of load buses and the set of renewable generators, respectively. The lower and upper bounds of the uniform distributions can be decided based on generation capacities and historical load data. The uniform distributions used to generate the training data are uncorrelated. Given a portfolio of stochastic grid variables (i.e., sample values of $P^L_i \; \forall i \in \mathcal{V}_L$ and $P^R_i \; \forall i \in \mathcal{V}_R$), (dispatch-able) power generation $P^T_i, \; \forall i \in \mathcal{V}_T$ and transmission line power flow $I_i, \; \forall i \in \mathcal{I}$ are uniquely determined by solving the DC OPF problem discussed in Section~\ref{sec:dcopf} ($\mathcal{V}_T$ and $\mathcal{I}$ represent the set of (dispatch-able) generators and transmission lines, respectively). The grid state data (generator dispatch and transmission line flows) thus obtained constitute the training data set.

Some of the QoIs needed for reliability and risk assessment, such as operating reserve and load shedding, are not directly available from the OPF solution. Operating reserve refers to the unused generation capacity that can be brought online in the specified time window, whereas load shedding refers to the amount of demand not met by the supply. For a given OPF solution, the operating reserve ($\Theta$) is taken as the residual (dispatch-able) power generation capacity after satisfying the net load. This can be computed as:
\begin{align}
    \Theta &= {\rm max} \left\{ \sum_{i \in \mathcal{V}_T} P_{i, max}^T - \left(\sum_{i \in \mathcal{V}_L} P_{i}^L - \sum_{i \in \mathcal{V}_R} P_i^R\right), \;\; 0 \right\}, 
\end{align}
where $P^T_{i, max}$ represents the maximum allowed generation capacity of dispatch-able generators. In case of load shedding, the supply is not able to meet the demand and the OPF solution cannot be obtained. Here, we use the slack bus to provide the required additional power. Power supply from the slack bus ($P_s$) is taken as load shedding ($\Psi$):
\begin{align}
    \Psi = \begin{cases}
              0, \;\;\;\; \text{if} \; P_s < 0 \\
              P_s, \;\;\; \text{if} \; P_s \geq 0.
          \end{cases}
\end{align}
The total cost, which is a system-level QoI, can also be calculated using the generator dispatch and the bid information.

\begin{figure*}[!ht]
     \centering
     \includegraphics[width=0.8\linewidth]{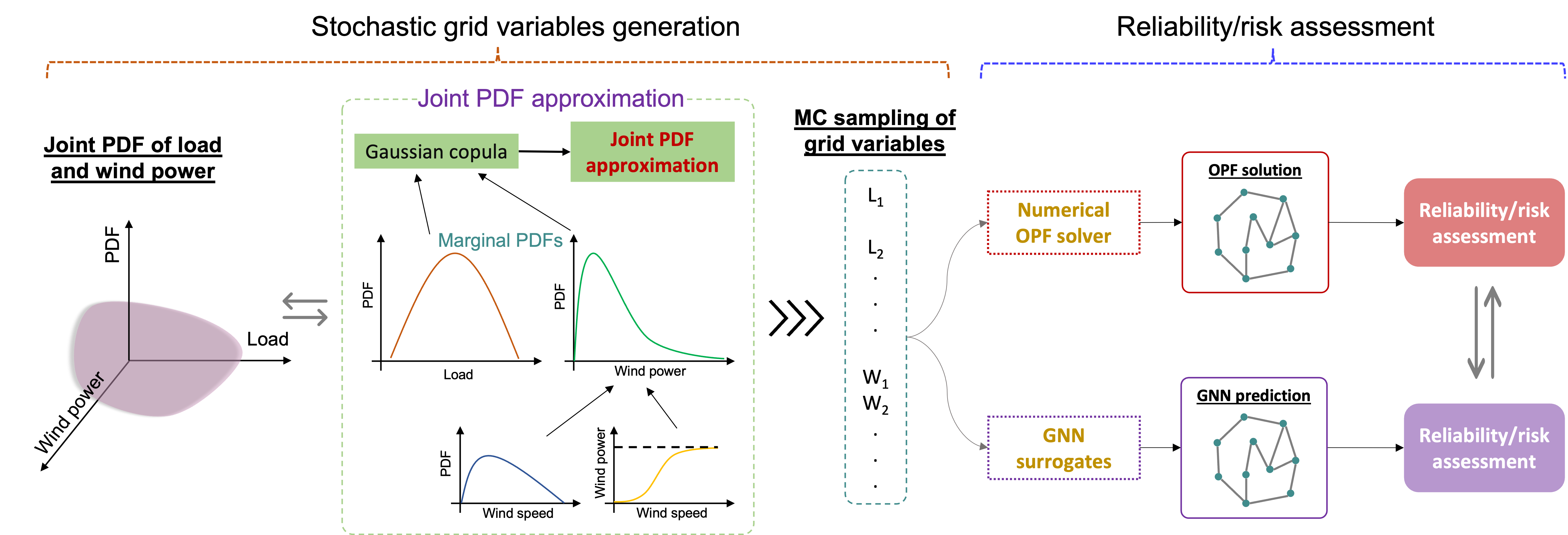}
     \caption{Schematic of test Dataset II generation and reliability and risk assessment}
     \label{fig:process_schematic}
\end{figure*}

\subsection{Testing data generation and GNN surrogate evaluation}
\label{sec:testing_data_generation}

Two testing datasets (I and II) are used to evaluate the GNN surrogate performance. Dataset I is generated using the same uncorrelated uniform distributions as the training data, but is \textit{separate and independent} of the training data. This type of test dataset has been used in the majority of the GNN surrogate literature in the power systems domain~\cite{Mohandes2019-mm,Zheng2015-gu}. Note, however, that the stochastic grid variables in a real power grid do not follow a uniform distribution. Previous studies suggest that the load follows a truncated normal distribution and different renewable generation sources are described by various other probability distributions~\cite{Hong2021-tn}. Wind power, for instance, is roughly proportional to the cube of wind speed, and wind speed is typically assumed to follow a Weibull distribution~\cite{Hong2021-tn}. In addition, grid variables are not independent but exhibit correlation~\cite{Li2007-ux,Aien2014-aq,Amor2014-ql}. The GNN surrogates are expected to handle such complicated yet more realistic statistical description (probability distributions and dependence structure) of the stochastic grid variables. Therefore, a second test dataset (Dataset II) is generated to test GNN surrogate performance with respect to this key requirement. 

To this end, the power grid is partitioned into multiple zones. Each zone contains a certain proportion of load buses, dispatch-able and RES generators. The joint PDF (forecast) of \emph{zonal} (aggregated) stochastic grid variables, i.e., load and RES power generation, is represented using a Gaussian copula~\cite{ tastu2013space} $G_C(\mathbf{u})$: $[0, 1]^{2s} \rightarrow [0, 1]$, for the sake of illustration, as: 
\begin{align}
    \label{eqn:Gaussian_copula}
    & G_C(\mathbf{u}) = \Phi_C\left(\Phi^{-1}(u_1), \Phi^{-1}(u_2),..., \Phi^{-1}(u_{2s})\right),
\end{align}
where $s$ is the number of distinct zones, $\mathbf{u} = [u_1, u_2, ..., u_{2s}]^T$ is the random vector containing independent uniform variables, $\Phi$ and $\Phi_C$ represent the standard normal CDF and multivariate normal CDF with zero mean and correlation matrix $C$, respectively. In this work, we simply assume a $C$ matrix and use the resulting copula for drawing samples of correlated grid variables ($\mathbf{u}$). Note that in the real-world application of GNN surrogates, samples drawn from the joint forecast PDF of stochastic variables will be used and will therefore correctly incorporate the correlations. Here, the Gaussian copula (with an assumed $C$ matrix) is used only for demonstrating and evaluating GNN-based reliability and risk quantification. Given a sample of $\mathbf{u}$, a sample of correlated stochastic grid variables can be generated as:
\begin{align}
\label{eqn:correlated_stochastic_variables}
S_m = \Phi_{S_m}^{-1}\left(G_C(\mathbf{u})\right), \; m=\{1, 2, ..., 2s\}, 
\end{align}
where $\Phi_{S_i}$ represents the marginal CDF of stochastic variable in zone $m$.

Without loss of generality, wind turbines are assumed to be the renewable generators in our illustration. Samples of zonal (aggregated) load ($L_i = S_m, m = 1, 2, ..., s$) and wind generation ($W_i = S_m, m = s+1, s+2, ..., 2s$) are drawn using the copula. The relative contribution of each load bus/wind turbine is assumed to be a constant in each zone. Consider zone $i$ containing $N_L$ load buses and $N_{W}$ wind turbines; then the bus-level load and wind power for zone $i$ are calculated as:
\begin{align} \label{eqn:relative_contribution}
\begin{split}
L_{i, j} &= r_{i, j}L_i, \; j = 1, 2, ..., N_L \\
W_{i, k} &= q_{i, k}W_i, \; k = 1, 2, ..., N_W    
\end{split}
\end{align}
where $r_{i, j}$ and $q_{i, k}$ represent the relative contribution of each load bus and wind generator bus to the respective zonal (aggregated) values, and $L_i$ and $W_i$ denote the aggregated values for zone $i$. Thus it is only necessary to draw samples of $L_i$ and $W_i$ for each zone, not at each bus. In this manner, a sample of stochastic grid variables with zonal correlation is obtained and used to solve the OPF problem. The procedure used to generate Dataset II is given below:
\begin{enumerate}[label=\roman*.]
    \item Draw a sample $L_i$ and $W_i$ for each zone using Eq.~\ref{eqn:correlated_stochastic_variables}
    \item Calculate load and wind power for load buses and wind generator buses in each zone using Eq.~\ref{eqn:relative_contribution}.
    \item Run a numerical solver \textit{Pandapower} to get OPF solution corresponding to the given sample of stochastic variables.
    \item Repeat Step i to iii multiple times to generate sufficient amount of test data samples for Dataset II.
\end{enumerate}

Note that Dataset II  enables the evaluation of GNN surrogate performance for reliability and risk estimation for realistic conditions (correlated grid variables with non-uniform probability distributions). The results from GNN surrogates-based reliability and risk assessment are compared with the ground truth (i.e., OPF-based reliability and risk assessment). The procedure for Dataset II generation and GNN-based reliability and risk assessment evaluation is illustrated in Fig.~\ref{fig:process_schematic}.

\subsection{Reliability and risk assessment}

The systematic risk assessment framework for grid-level QoIs proposed by Stover et al.~\cite{Stover2023-ov} is employed for evaluating and communicating the hours-ahead operational risk. Specifically, risk is defined by a risk triplet: a credible adverse event, the likelihood of the event occurring and the consequence of its occurrence. Level 2 and 3 metrics defined in~\cite{Stover2023-ov} are considered in this work. Level 2 metrics quantify the probability of failure corresponding to a chosen failure mode (reserve inadequacy, loss of load, etc.). Level 3 metrics consider the consequence cost of the adverse event in addition to the probability of failure. We consider reserve inadequacy as the failure mode for zonal and system level reliability and risk assessment. At the transmission line level, we consider flow beyond a certain threshold as the failure event. This event is a surrogate for transmission line overloading.

\subsubsection{Reserve adequacy}

To ascertain operating reserve adequacy, we use the minimum reserve requirement (MRR) as the reliability threshold. MRR is typically set as the generation capacity of the single largest generator in the power grid~\cite{Stover2023-ov}. Therefore, the probability of reserve inadequacy is given by: 
\begin{align}
    \label{system_reliability}
    \mathcal{P}^S_f(\Phi, \text{MRR}) = p(\Phi < \text{MRR}),
\end{align}
and is computed as 
\begin{align}
    \label{system_reliability_comp}
    \mathcal{P}^S_f(\Phi, \text{MRR}) = \frac{1}{M}\sum_{k=1}^M V(k),
\end{align}
where $M$ denotes the number of samples, $V(k) = 1$ if $\Phi < \text{MRR}$ holds for $k$-th sample and $V(k) = 0$ otherwise. Note that MRRs at system and zone levels are considered in this work. 

Next, risk is computed as the product of probability of failure and the consequence (monetary cost). A constant consequence cost, $\mathcal{C}^S$, is assumed for insufficient reserve; thus the corresponding risk can be calculated as:
\begin{align}
    \label{system_risk}
    \Re^S(\Phi) = \mathcal{P}^S_f(\Phi, \text{MRR}) \mathcal{C}^S.
\end{align}

\subsubsection{Branch overloading}

The risk assessment framework developed by Stover et al.~\cite{Stover2023-ov} is extended to include reliability and risk quantification for transmission line flow. This extension can help grid operators identify the risk of congestion in the power grid. The electric current $l_f$ in one or more branches exceeding a certain threshold $\epsilon$ is considered as a failure event for this purpose. The probability of overloading branch $i$ is defined as:
\begin{align}
    \label{eqn:individual_branch_reliability}
    \mathcal{P}^B_f(i) &= p(l_{f, i} \geq \epsilon),
\end{align}
and computed as
\begin{align}
    \label{eqn:individual_branch_reliability_comp}
    \mathcal{P}^B_f(i) & = \frac{1}{M}\sum^M_{k=1} U_k(i),
\end{align}
where $i = \{1, \;2, \; \dots, \; N\}$ with $N$ denoting the number of branches. $U_k(i)=1$ if $l_f \geq \epsilon$ holds for branch $i$ in the $k$-th sample, otherwise 0. Moreover, the conditional probability of overloading a transmission line $j$, given that transmission line $i$ is overloaded, is also considered. This probability is defined and computed as:
\begin{align}
    \label{eqn:multiple_branches_reliability}
    \mathcal{P}^B_f(j|i) &= p(l_{f, j} | \Omega_i) = \frac{1}{|\Omega_i|}\sum_{k\in \Omega_i}U_k(j),
\end{align}
where $j = \{1, \;2, \; \dots, \; N\}$ and $j\neq i$. $\Omega_i$ represents the set of samples with $l_{f, i} \geq \epsilon$.

The overall risk of branch overloading, given that branch $i$ is overloaded, can computed by combining Eq.~\ref{eqn:individual_branch_reliability} and~\ref{eqn:multiple_branches_reliability} as:
\begin{align}
    \label{branch_risk}
    \Re^B(i) = \mathcal{P}^B_f(i) \mathcal{C}_i^B + \sum^N_{j=1, \; j\neq i}\mathcal{P}^B_f(j|i)\mathcal{C}_j^B.
\end{align}
where $\mathcal{C}_i^B$ and $\mathcal{C}_j^B$ are the consequence costs of overloading of lines $i$ and $j$ respectively. This formula includes the probability of overloading branch $i$ and the conditional probability of overloading any other branch, given that branch $i$ is overloaded; thus it provides a meaningful estimate of \textit{overall} branch overloading risk. This overall risk can be used by operators to identify the most critical branches in the grid, given a probabilistic forecast of stochastic grid variables.

In summary, this section outlines the procedure of using GNN surrogates to estimate power grid operational reliability and risk. This includes GNN surrogate development, training/test data generation and reliability/ risk quantification. A system-level reliability and risk assessment framework is extended to branch level to account for line flow constraints. The performance of GNN surrogates is evaluated based on both simple (uniform) and complex (realistic) probabilistic forecasts of grid variables.

\section{Numerical Experiments}
\label{sec:numerical_experiment}

The proposed GNN surrogate-based risk quantification method is evaluated on four benchmark power grids: Case118~\cite{zimmerman1997matpower}, Case300~\cite{pan2020deepopf}, Case1354pegase~\cite{fliscounakis2013contingency} and Case2848rte~\cite{josz2016ac}. Without loss of generality, we consider wind power as renewable source and approximately 20\% generators are assumed to be wind turbines. Due to the low cost and environmental friendliness, renewable energy sources, whenever available, would preferably be fully deployed to meet demand in real-world application. To enforce this, we always use the wind generator output first towards meeting the demand, and other sources are deployed to satisfy the remaining demand.

For training data and testing Dataset I, bus-level load and wind power generation are randomly drawn from independent and identical uniform distributions. A sample of the stochastic variables is given as input to the numerical solver \emph{Pandapower} to obtain the OPF solution (bus-level, branch-level and system-level QoIs). A total of 1000 samples (70\% for training and 30\% for Dataset I) are generated using the OPF solver. Although these samples share the same grid topology, they are distinct graphs because feature vector of each node is different. Supervised learning is used during the training process and the hyper-parameters are carefully tuned to promote model performance.

For the synthesis of Dataset II, the grids are partitioned into zones, and each zone contains a specified number of load buses and generators (both renewable and other dispatchable generators). Following the procedure outlined in Section~\ref{sec:testing_data_generation}, the joint PDF approximation is established using marginal PDFs via Eq.~\ref{eqn:Gaussian_copula} and~\ref{eqn:correlated_stochastic_variables}. We consider demand following a truncated normal (TN) distribution, while wind power generation is converted from wind speed which is assumed to follow a Weibull (WB) distribution. The conversion follows the power rating curve in~\cite{Hong2021-tn}. Details of zonal partitioning of the four grids, zonal correlations, the parameters used to define the marginal distributions, the training loss versus the number of training samples, as well as other important details of the numerical experiments can be found in the Supplementary Information (SI) provided with this article. Note that the training loss behavior suggests that it is not necessary to use more samples in the training process.

\section{Results}
\label{sec:results}

The prediction accuracy of GNN surrogates for bus-level, branch-level and system-level QoIs is quantified using relative error with respect to a reference solution, which is obtained using numerical OPF solver \textit{Pandapower}~\cite{thurner2018pandapower}. The accuracy of GNN-based reliability and risk quantification w.r.t. reserve adequacy at the system and zonal level is examined, while twenty critical branches are selected to demonstrate the results of branch overloading analysis.

\subsection{GNN surrogate prediction accuracy}

\begin{table}[!ht]
    \centering
    \caption{\textsc{Maximum relative error (\%) of gnn prediction at different magnitudes of reference solution}}
    \label{tab:GNN_performance_detail}
    \begin{tabular}{ccccccccc}
        \toprule
        Magnitude ($\rightarrow$)& \multicolumn{2}{c}{[0.1, 1)} & \multicolumn{2}{c}{[1, 10)} & \multicolumn{2}{c}{[10, 100)} & \multicolumn{2}{c}{[100, 300)} \\
       (MW)& & & & \\
        \cmidrule{2-9}
        Grids ($\downarrow$) & PG & PF & PG & PF & PG & PF & PG & PF \\
        \midrule
        Case118  & 25.5 & 40.3 & 2.3 & 3   & 1.5 & 1.7 & 0.5 & 0.6 \\
        Case300  & 24.2 & 38.7 & 2.1 & 2.5 & 1.8 & 1.5 & 0.3 & 0.4 \\
        Case1354 & 23.7 & 42.5 & 2.6 & 3   & 2   & 2.1 & 1   & 0.8 \\
        Case2848 & 26.1 & 36.2 & 2.2 & 2.7 & 2.5 & 2.3 & 1.3 & 1.5 \\
        \bottomrule
    \end{tabular}
\end{table}

We first evaluate GNN prediction for bus-level (power generation, or PG) and branch-level (power flow, or PF) QoIs. The results are shown in Table~\ref{tab:GNN_performance_detail}. In Table~\ref{tab:GNN_performance_detail}, the maximum relative percentage error is reported for four magnitudes of reference (true, or OPF-based) values, i.e., [0.1, 1), [1, 10), [10, 100), and [100, 300] MW. For power generation and flow magnitude ranging between 0.1 and 1,  the maximum prediction error are $26\%$ and $42\%$ for active power and transmission line flow prediction, respectively. This is because a small magnitude of reference value results in high relative error; however, note that these correspond to small absolute error. Such errors are not likely to cause constraint violation and significant discrepancy in the objective (cost) function. With regard to the other three categories, maximum prediction error drops below $3\%$. Taking the active power prediction in Case2848 as example, the maximum error declines to $2.5\%$ when the magnitude is in the range [10, 100), and further drops to $1.3\%$ in the range [100, 300). Similar results are observed for other grids. These maximum percentage errors demonstrate that GNN can provide accurate prediction for bus-level and branch-level predictions.

System-level prediction results are summarized in Table~\ref{tab:GNN_performance}. The mean relative error in reserve prediction is $\sim 0.22 \%$ for all grids. Even lower error ($<0.2\%$) is observed for load shedding prediction. The average error in cost prediction ranges from $0.2\%$ to $0.4\%$ for different grids. For comparison, results from recent DC OPF surrogate modeling investigations are also shown in Table~\ref{tab:GNN_performance}. Slightly better accuracy has been reported in~\cite{wu2022fast}~and~\cite{pan2020deepopf}. However, these previous studies have not reported results for the larger grids, Case1354 and Case2848. Our work is mainly concerned with risk quantification and not with decision-making, hence the small errors in cost prediction for small grids are acceptable. GNN surrogates demonstrate consistently good performance for grids of different sizes. This is particularly important for practical usage. Overall the results suggest that GNN surrogates are capable of predicting system-level QoIs with excellent accuracy.

The efficiency (speed up) achieved using GNN surrogates compared to the original OPF model is also shown in Table~\ref{tab:GNN_performance}. For small grids (i.e., Case118 and Case300), GNN surrogates are 250 and 300 times faster than numerical DC OPF solver. Similar speedup has been reported by previous studies~\cite{wu2022fast,pan2020deepopf}. For the two large grids, the GNN surrogate model achieves more than 500 times speed up compared to the original PDF. These results encourage investigation regarding the accuracy of reliability/risk estimates obtained using GNN surrogates. If this speed up can support sufficiently accurate reliability/risk estimation, then the proposed risk quantification methodology could be used by grid operators for real-time decision-making.
\begin{table*}[!ht]
    \centering
    \caption{\textsc{gnn surrogate performance}}
    \label{tab:GNN_performance}
    \begin{tabular}{ccccccccccc}
        \toprule
        \multirow{2}{*}{Grid} & \multicolumn{2}{c}{Mean relative \% error} & \multicolumn{4}{c}{Mean relative \% error, Cost} & \multicolumn{4}{c}{Speed up} \\ 
        \cmidrule{4-11} 
        & Reserve & Shedding & \cite{pan2020deepopf} & \cite{zhao2020deepopf+} & \cite{wu2022fast} & This work & \cite{pan2020deepopf} & \cite{zhao2020deepopf+} & \cite{wu2022fast} & This work \\
        \midrule
        Case118 & 0.20 & 0.15 & 0.2 & 0.56 & 0.1 & 0.2 & $\times$281 & $\times$208 & $\times$350 & $\times$250 \\
        Case300 & 0.20 & 0.17 & 0.1 & 0.66 & 0.2 & 0.3 & $\times$318 & $\times$126 & $\times$200 & $\times$300 \\
        Case1354 & 0.21 & 0.17 & $-$ & $-$ & $-$ & 0.3 & $-$ & $-$ & $-$ & $\times$500 \\ 
        Case2848 & 0.22 & 0.18 & $-$ & $-$ & $-$ & 0.4 & $-$ & $-$ & $-$ & $\times$800 \\
        \bottomrule
    \end{tabular}
\end{table*}

\subsection{GNN surrogates for reliability and risk assessment}

\subsubsection{Reserve adequacy}

Reliability assessment results for system-level reserve adequacy are shown in Table~\ref{tab:system_level_reliability}. The GNN surrogate-based prediction of probability of inadequate reserve demonstrates excellent accuracy. Taking Case118 as example, zone III is the most vulnerable to reserve shortage with $\mathcal{P}^s_f = 0.21$, followed by zones II and I at 0.16 and 0.12. The corresponding GNN-based predictions are 0.20, 0.11 and 0.16, respectively. Risk assessment results are reported in Table~\ref{tab:system_level_risk}. GNN-based risk estimates also show remarkable agreement with the OPF-based risk estimate. The reliability/risk estimation accuracy and speed up (Table~\ref{tab:GNN_performance}) indicate that GNN surrogates can enable MC sampling-based real-time operational risk estimation by quickly providing thousands of (approximate) OPF solutions, given the updated forecast of the stochastic grid variables.
\begin{table*}[!ht]
    \centering
    \caption{\textsc{Probability of reserve inadequacy at different zones ($\ast$: opf, $\circ$: gnn)}}
    \label{tab:system_level_reliability}
    \begin{tabular}{cccccccccccccccccc}
        \toprule
        \multicolumn{2}{c}{Grid} & I & II & III & IV & V & VI & VII & VIII & IX & X & XI & XII & XIII & XIV & XV & XVI \\
        \midrule
        \multirow{2}{*}{Case118} & $\ast$ & 0.12 & 0.16 & 0.21 \\
        & $\circ$ & 0.11 & 0.16 & 0.20 \\
        \multirow{2}{*}{Case300} & $\ast$ & 0.15 & 0.74 & 0.26 & 0.53 \\
         & $\circ$ & 0.14 & 0.75 & 0.27 & 0.53 \\
        \multirow{2}{*}{Case1354} & $\ast$ & 0.85 & 0.18 & 0.05 & 0.36 & 0.28 & 0.53 & 0.31 & 0.30 \\
        & $\circ$ & 0.84 & 0.17 & 0.06 & 0.36 & 0.28 & 0.54 & 0.31 & 0.30 \\
        \multirow{2}{*}{Case2848} & $\ast$ & 0.59 & 0.90 & 0.89 & 0.82 & 0.04 & 0.69 & 0.38 & 0.52 & 0.66 & 0.19 & 0.27 & 0.72 & 0.78 & 0.85 & 0.78 & 0.04 \\
        & $\circ$ & 0.58 & 0.90 & 0.89 & 0.81 & 0.04 & 0.70 & 0.38 & 0.51 & 0.66 & 0.19 & 0.28 & 0.73 & 0.78 & 0.85 & 0.78 & 0.04 \\
        \bottomrule
    \end{tabular}
\end{table*}

\begin{table*}[!ht]
    \centering
    \caption{\textsc{Risk of reserve inadequacy at different zones ($\ast$: opf, $\circ$: gnn), $\times 10^3$}}
    \label{tab:system_level_risk}
    \begin{tabular}{cccccccccccccccccc}
        \toprule
        \multicolumn{2}{c}{Grid} & I & II & III & IV & V & VI & VII & VIII & IX & X & XI & XII & XIII & XIV & XV & XVI \\
        \midrule
        \multirow{2}{*}{Case118} & $\ast$ & 1.10 & 1.58 & 2.44 \\
        & $\circ$ & 1.04 & 1.49 & 2.40 \\
        \multirow{2}{*}{Case300} & $\ast$ & 2.15 & 2.87 & 4.16 & 2.06 \\
        & $\circ$ & 2.09 & 2.78 & 4.12 & 2.09 \\
        \multirow{2}{*}{Case1354} & $\ast$ & 23.5 & 10.9 & 16.9 & 22.9 & 15.7 & 29.0 & 13.2 & 17.8 \\
        & $\circ$ & 23.9 & 10.9 & 15.6 & 21.8 & 16.3 & 28.0 & 12.8 & 16.0 \\
        \multirow{2}{*}{Case2848} & $\ast$ & 47.6 & 56.9 & 56.7 & 54.5 & 31.1 & 50.8 & 41.4 & 45.6 & 49.7 & 35.8 & 38.2 & 51.6 & 53.5 & 55.5 & 53.3 & 31.1 \\
        & $\circ$ & 46.1 & 57.9 & 55.7 & 53.5 & 32.5 & 52.6 & 41.6 & 44.3 & 50.8 & 35.8 & 38.7 & 53.0 & 53.3 & 55.5 & 54.0 & 31.7 \\
        \bottomrule
    \end{tabular}
\end{table*}

\subsubsection{Branch overloading}

\begin{figure*}[!ht]
    \centering
    \includegraphics[width=0.8\linewidth]{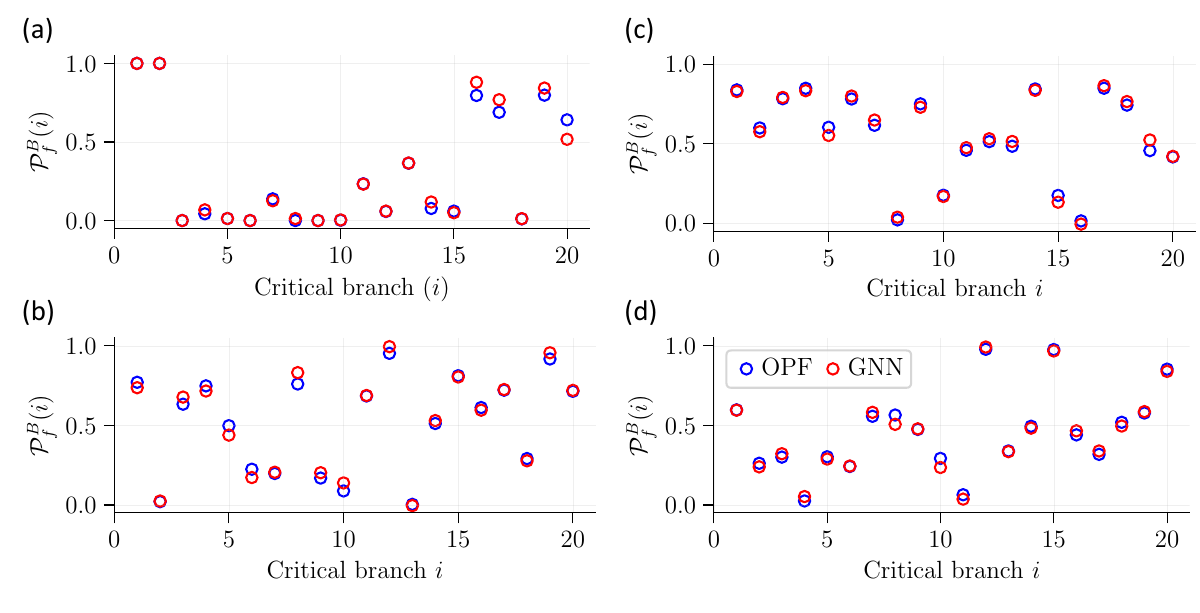}
    \caption{Reliability assessment at branch level, twenty critical branches are selected to evaluate the GNN surrogate accuracy. (a)-(d): Case118, Case300, Case1354 and Case2848, respectively.}
    \label{fig:branch_overloading_reliability}
\end{figure*}

\begin{figure*}[!ht]
    \centering
    \includegraphics[width=0.8\linewidth]{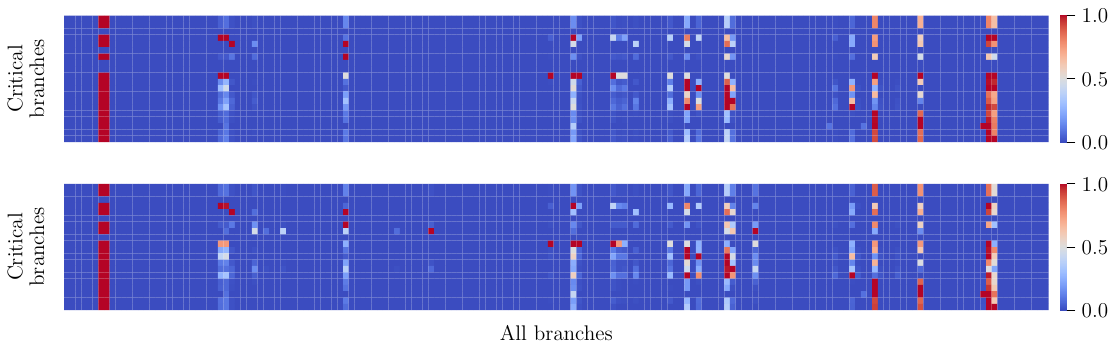}
    \caption{Conditional probability (Eq.~\ref{eqn:multiple_branches_reliability}) of branch overloading (Case118). Top: OPF, bottom: GNN.}
    \label{fig:branch_overloading_cond_reliability}
\end{figure*}

Results of branch overloading analysis are shown in Fig.~\ref{fig:branch_overloading_reliability}. Overall, GNN surrogates provide accurate prediction for the probability of overloading in individual transmission line. Herein we provide detailed analysis for the results of Case118. It is noticed that there are two branches with $\mathcal{P}^B_f(i) = 1$, indicating they will always approach the maximum allowed capacity and thus operators should give special attention to these lines to ensure the safety of grid operation. There are four branches with $\mathcal{P}^B_f(i) > 0.5$, and they have been identified as such by the GNN surrogate-based assessment. The probability of multiple branch overloading is shown in Fig.~\ref{fig:branch_overloading_cond_reliability}. Given a heavily loaded critical branch, the conditional probability of overloading all the remaining branches is calculated. Overall, the GNN predictions are in good agreement with the ground truth. It is noticed that overloading does not happen for most branches as the value of $\mathcal{P}^B_f(j|i)$ is almost zero. Moreover, there are a few branches (e.g., $j = {6, 7, 89, 142, 150, 162, 163}$) that experience heavier loading more often than other branches. It is therefore necessary to monitor the flow in these lines during daily operation of the grid. 

\begin{figure*}[!ht]
    \centering
    \includegraphics[width=0.8\linewidth]{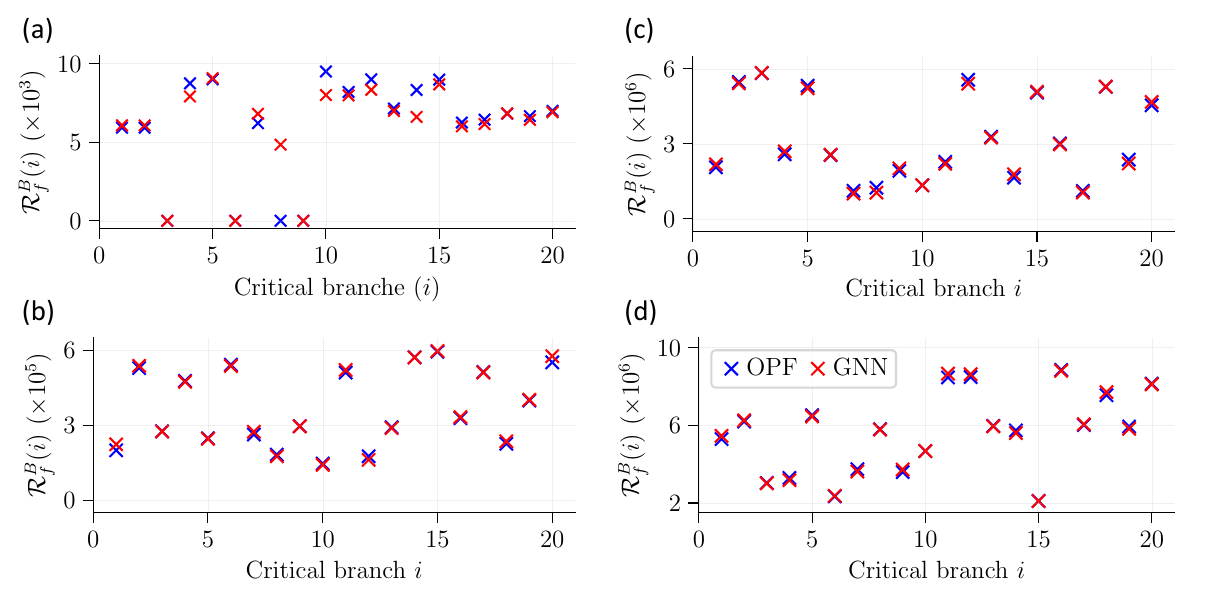}
    \caption{Risk assessment at branch level, twenty critical branches are selected to evaluate the GNN surrogate accuracy. (a)-(d): Case118, Case300, Case1354 and Case2848, respectively.}
    \label{fig:branch_overloading_risk}
\end{figure*}

The overall risk of branch overloading, as defined in Eq.~\ref{eqn:multiple_branches_reliability}, is also shown in Fig.~\ref{fig:branch_overloading_risk}. It is observed that $\Re^B(i)$ ranges between $\$0$ and $\$10,000$. Interestingly, it is found that branches with the highest (marginal) overloading probability do not carry  the highest overall (system-wide) overloading risk. This indicates that there will be significant risk of overall (system-wide) branch overloading if the branches that rarely experience overloading experience high power flow. The comparison of OPF-based and GNN-based risk estimates indicates that GNN surrogates are sufficiently accurate for predicting this complex (conditional) risk, and can be used for real-time risk assessment.

\subsubsection{GNN surrogate accuracy for different grid forecasts}

\begin{figure}
    \centering
    \includegraphics[width=0.9\linewidth]{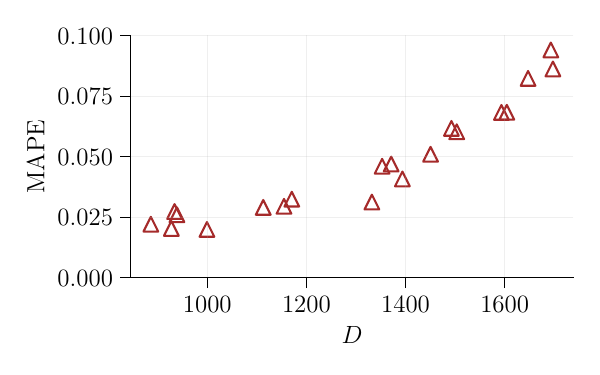}
    \caption{Effect of distance between training and testing data on GNN prediction accuracy. The results are based on Case118 grid.}
    \label{fig:energy_distance}
\end{figure}

The GNN surrogate trained using the training data  with simple underlying probability distributions (uniform, uncorrelated variables) needs to be exercised for grid variable forecasts that follow more realistic probability distributions (non-uniform, correlated variables) for reliability and risk assessment. It is thus important to quantify the error in reliability or risk estimation as a function of a measure of distance between the training and the forecast distributions. To this end, multiple joint forecast distributions are generated by modifying the marginal distribution parameters. Samples of correlated grid variables are obtained from each of these distributions and are used to quantify reliability and risk. The similarity between the training data and a candidate (forecast) supply/demand regime (as represented by the corresponding joint probability distributions) is quantified using a statistical distance metric ($D$). If the samples of a random vector ${X}$ are given by $x_1, x_2, \dots, x_n$ and those of a random vector $Y$ are given by $y_1, y_2, \dots, y_m$, then $D$ is defined as:
\begin{align}
    \label{eqn:statistical_distance}
    D = E [||X - Y||] =\frac{1}{mn}\sum_{i=1}^n\sum_{j=1}^m||x_i - y_j||,
\end{align}
where $n$ and $m$ are the number of samples of $X$ and $Y$, respectively, $E$ denotes the expected value, and $||\bullet||$ is a vector norm. A higher value of $D$ indicates a higher degree of dissimilarity between $X$ and $Y$. In the context of GNN surrogate evaluation, $X$ represents training data whereas $Y$ denotes testing data. The (ensemble-based) distance ($D$) between GNN training data and the forecast distribution is computed using Eq.~\ref{eqn:statistical_distance}. The mean absolute percentage error (MAPE) in the estimation of branch overloading probability for different values of $D$ is shown in Fig.~\ref{fig:energy_distance}. The MAPE $\leq 5$\% when $D < 1200$, however, the value of MAPE gradually increases after that till it reaches about $10$\% for $D = 1600$. The methodology used in this exercise could be implemented by grid operators to quantify the error bounds on their GNN-based real-time risk estimates.

\section{Conclusion}
\label{sec:conclusion}

We investigated the utility of DC OPF surrogate modeling for power grid operational risk assessment. We developed a methodology for evaluating such \emph{risk surrogates} and demonstrated the methodology for GNN surrogates. To this end, we solved a large number of optimal power flow (OPF) problems to obtain the surrogate model training data, and learned multiple GNN surrogates that predict bus-level, branch-level and system-level QoIs. We considered inadequate zonal and system reserve, as well as branch overloading as the relevant failure modes for reliability and risk assessment. The surrogate model performance for predicting the QoI values as well as for quantifying the reliability and risk was assessed using two types of test datasets. The first dataset used simple statistical description (uniform, uncorrelated) for stochastic grid variables (renewable generation and load). This is the prevailing practice in literature on OPF surrogate models. The second dataset used realistic, complex statistical description of stochastic grid variables. Specifically, we used realistic marginal probability distributions and employed a Gaussian copula to capture the dependence structure of stochastic grid variables. This has not been previously reported in the power grid literature, and GNN surrogate performance evaluation under this complicated joint (forecast) distribution is needed before real-world operational risk assessment of power grids. We demonstrated the proposed GNN surrogate-based reliability and risk estimation methodology on the Case118, Case300, Case1354pegase and Case2848rte power grids. By comparing OPF-based and GNN-based reliability and risk estimation, we showed that GNN surrogates can provide accurate estimate of the grid's operational risk at a desired time instant and are also capable of dealing with correlations between variables in distinct zones. We also developed a methodology that could be used to estimate the error in surrogate-based reliability and risk estimation as a function of the statistical distance between training data and grid forecast data. The excellent accuracy and lower computational cost of the GNN surrogates as compared to numerical OPF solvers indicate that GNNs can be good surrogates for computationally expensive numerical OPF solvers, and could be deployed for real-time risk assessment. 

The proposed methodology has only considered static reliability and risk assessment, for a given unit commitment. It can be extended to capture temporal evolution of grid variables. To that end, it is necessary to examine generator on/off status, ramping rate, wear and tear cost, etc., and consider the change in committed generator units. In addition, advanced learning methods such as reinforcement learning maybe helpful to improve GNN prediction accuracy, especially for branch-level and system-level QoIs, and thus enhance real-time reliability and risk quantification. 

\section{Acknowledgement}
\label{sec:acknowledgement}

This work is partly funded by ARPA-E PERFORM award DE-AR0001280. The support is gratefully acknowledged.

\bibliographystyle{IEEEtran}
\bibliography{references}

\end{document}